\documentclass[runningheads]{llncs}

\usepackage[utf8]{inputenc}
\usepackage[T1]{fontenc}
\usepackage{graphicx}
\usepackage{amsmath}
\usepackage{booktabs}
\usepackage{hyperref}
\usepackage{xcolor}
\usepackage{listings}
\usepackage{subcaption}
\usepackage{float}

\begin{document}

\title{Structured Linked Data as a Memory Layer\\
for Agent-Orchestrated Retrieval}

\titlerunning{Structured Linked Data for Agent-Orchestrated Retrieval}

\author{Andrea Volpini \and Elie Raad \and Beatrice Gamba \and David Riccitelli}

\authorrunning{Volpini et al.}

\institute{WordLift, Rome, Italy\\
\email{\{andrea, elie, beatrice, david\}@wordlift.io}}

\maketitle

\begin{abstract}
Retrieval-Augmented Generation (RAG) systems typically treat documents as
flat text, ignoring the structured metadata and linked relationships that
knowledge graphs provide. In this paper, we investigate whether structured
linked data---specifically Schema.org markup and dereferenceable entity
pages served by a Linked Data Platform---can improve retrieval accuracy
and answer quality in both standard and agentic RAG systems.

We conduct a controlled experiment across four domains (editorial, legal,
travel, e-commerce) using Vertex AI Vector Search 2.0 for retrieval and
the Google Agent Development Kit (ADK) for agentic reasoning. Our
experimental design tests seven conditions: three document representations
(plain HTML, HTML with JSON-LD, and an enhanced agentic-optimized entity
page) crossed with two retrieval modes (standard RAG and agentic RAG
with multi-hop link traversal), plus an Enhanced+ condition that adds
rich navigational affordances and entity interlinking.

Our results reveal that while JSON-LD markup alone provides only modest
improvements ($\Delta = +0.17$, $p_{\text{adj}} = 0.024$), our enhanced entity page
format---incorporating \texttt{llms.txt}-style agent instructions,
breadcrumbs, and neural search capabilities---achieves substantial
gains: +29.6\% accuracy improvement for standard RAG ($p < 10^{-21}$,
$d = 0.60$) and +29.8\% for the full agentic pipeline ($p < 10^{-21}$,
$d = 0.61$). The Enhanced+ variant, with richer navigational
affordances, achieves the highest absolute scores (accuracy: 4.85/5,
completeness: 4.55/5) though the incremental gain over the base
enhanced format is not statistically significant ($d = 0.08$).
We release our dataset, evaluation framework,
and enhanced entity page templates to support reproducibility.

\end{abstract}

\section{Introduction}
\label{sec:intro}

The rise of Generative AI has fundamentally changed how users access
information online. Search engines increasingly augment traditional
results with AI-generated summaries---a paradigm exemplified by
Google's AI Mode, which retrieves, reasons over, and synthesizes
information from multiple web sources. Understanding and optimizing
for this new retrieval paradigm is critical for content creators,
marketers, and organizations that depend on search visibility.

Retrieval-Augmented Generation (RAG) has emerged as the dominant
architecture for grounding large language model (LLM) outputs in
factual, up-to-date information~\cite{lewis2020rag}. However, most
RAG implementations treat documents as unstructured text, discarding
the rich structured metadata that many websites already provide via
Schema.org markup and knowledge graph representations.

In this paper, we ask: \textbf{Can structured linked data improve
RAG performance, and does agentic link traversal unlock further gains?}

Our work is motivated by three observations:
\begin{enumerate}
    \item Websites increasingly embed Schema.org JSON-LD structured data,
          yet RAG systems rarely exploit this metadata.
    \item Linked Data Platforms serve entity pages that support content
          negotiation, enabling programmatic traversal of knowledge graphs.
    \item Agentic AI systems (those capable of planning, tool use, and
          multi-step reasoning) can follow links and aggregate information
          across entity boundaries---mimicking the behavior of AI-powered
          search engines.
\end{enumerate}

We make the following contributions:
\begin{itemize}
    \item A controlled experimental framework comparing seven conditions
          (3 document formats $\times$ 2 retrieval modes + an Enhanced+
          variant) across four industry verticals, with 2,443 individual
          query evaluations.
    \item An \emph{enhanced entity page} format designed to maximize
          both human readability and agentic discoverability, incorporating
          \texttt{llms.txt}-style instructions and neural search capabilities,
          and an Enhanced+ variant with richer navigational affordances.
    \item Empirical evidence showing that enhanced entity pages yield
          the strongest improvements: +29.6\% accuracy in standard RAG
          ($d = 0.60$) and +29.8\% in the agentic pipeline
          ($d = 0.61$), while JSON-LD markup alone provides only
          marginal improvements.  The Enhanced+ variant achieves the
          highest absolute scores but offers no statistically significant
          gain over the base enhanced format.
    \item A reusable dataset and evaluation harness released for
          reproducibility.
\end{itemize}

\section{Related Work}
\label{sec:related}

\subsection{Generative Engine Optimization}

Aggarwal et al.~\cite{aggarwal2023geo} introduced Generative Engine
Optimization (GEO), demonstrating that content optimization strategies
such as adding citations, statistics, and authoritative language can
boost visibility in generative search engines by up to 40\%.
Our work extends GEO from \emph{visibility optimization} to
\emph{retrieval accuracy}, focusing on structured data as the
optimization lever.

\subsection{Retrieval-Augmented Generation}

RAG was formalized by Lewis et al.~\cite{lewis2020rag}, who combined
a pre-trained sequence-to-sequence model with a dense retriever to
ground generation in retrieved passages. Subsequent work explored
pre-training with retrieval objectives~\cite{guu2020realm} and
scaling retrieval corpora to trillions of tokens~\cite{borgeaud2022retro}.
More recently, Self-RAG~\cite{asai2024selfrag} introduced self-reflection
mechanisms for adaptive retrieval, enabling models to decide when and
what to retrieve. Trivedi et al.~\cite{trivedi2023interleaving}
demonstrated that interleaving retrieval with chain-of-thought reasoning
significantly improves multi-step question answering.

Despite these advances, existing RAG systems predominantly operate on
unstructured text. Our work bridges this gap by demonstrating that
structured metadata---specifically Schema.org JSON-LD---provides
a complementary signal that improves retrieval quality.

\subsection{Knowledge Graphs and Structured Data on the Web}

The vision of a machine-readable web was articulated by
Berners-Lee et al.~\cite{berners2001semantic} and operationalized
through Linked Data principles~\cite{bizer2009linked}. Schema.org,
launched in 2011 as a collaboration among major search engines,
provides a shared vocabulary for structured data on the
web~\cite{guha2016schemaorg,schemaorg}. Today, over 40\% of
web pages include Schema.org markup~\cite{guha2016schemaorg}.

Knowledge graphs have become central to both academic research and
industry applications~\cite{hogan2021knowledge,noy2019industry}.
Early efforts to bring structured data to content management systems
include WordLift~\cite{volpini2015wordlift}, which introduced semantic
annotation and entity-based navigation for WordPress sites, and
MICO~\cite{aichroth2015mico}, which developed linked-data pipelines
for multimedia content enrichment.
Recent surveys examine the unification of LLMs and knowledge
graphs~\cite{pan2024graphrag}, while Graph RAG approaches
explicitly leverage graph structure during
retrieval~\cite{peng2024graphretrieval}.

Several recent systems construct retrieval graphs from documents to
improve RAG.  LightRAG~\cite{guo2024lightrag} builds a graph index
from document-extracted entities and relations, using dual-level
retrieval (low-level for specific facts, high-level for topics) to
outperform traditional RAG on multi-hop questions.
HippoRAG~\cite{gutierrez2024hipporag} models retrieval after the
hippocampal memory indexing theory, constructing a knowledge graph
from passages and using personalized PageRank for context-sensitive
retrieval.  Both systems demonstrate the value of graph structure
for retrieval, but differ from our approach in a fundamental way:
they construct purpose-built graphs at indexing time from unstructured
text, whereas we leverage \emph{existing} structured data already
published on the web via Schema.org and Linked Data Platforms.  Our
approach requires no graph construction step---the knowledge graph
is the publisher's source of truth, maintained independently of the
retrieval system, and accessible through dereferenceable URIs that
support content negotiation.

\subsection{Agentic AI and Tool-Augmented LLMs}

Agentic AI systems extend LLMs with the ability to plan, use tools,
and reason over multiple steps. Yao et al.~\cite{yao2023react}
introduced ReAct, interleaving reasoning traces with action steps.
Schick et al.~\cite{schick2023toolformer} demonstrated that LLMs
can learn to use external tools autonomously. The Google Agent
Development Kit (ADK)~\cite{googleadk2025} provides a production
framework for building multi-tool agents.

Multi-hop question answering~\cite{mavi2024multihop}---where
answering requires combining information from multiple
sources---is a natural application for agentic systems. The
Model Context Protocol (MCP)~\cite{mcp2024} provides a standardized
interface for LLM--tool integration. Our
agentic RAG configuration enables link traversal across entity
boundaries, effectively mimicking the behavior of AI-powered
search systems that follow links to aggregate information.

\section{Methodology}
\label{sec:method}

\subsection{System Architecture and AI Mode Parallel}

Our experimental system mirrors the emerging architecture of AI-powered
search engines such as Google's AI Mode, which retrieves web pages,
reasons over their structured content, and synthesizes multi-source
answers.  Our pipeline reproduces this pattern using Google
Cloud components for retrieval and reasoning, combined with an
independent knowledge graph for structured data:

\begin{itemize}
    \item \textbf{Vertex AI Vector Search 2.0}~\cite{vertexai2025} serves
          as the retrieval backbone.  Unlike traditional vector databases,
          Vector Search 2.0 is designed as a self-tuning, fully managed,
          AI-native search engine.  It combines dense semantic search
          (via \texttt{text-embedding-005} embeddings) with sparse keyword
          matching in a single hybrid query, automatically tuning retrieval
          parameters.  This mirrors how AI Mode identifies candidate web
          pages from a large corpus.
    \item \textbf{Google Agent Development Kit (ADK)}~\cite{googleadk2025} powers
          the agentic reasoning layer, providing a ReAct-style
          loop~\cite{yao2023react} with tool-use capabilities.  Like AI
          Mode's multi-step reasoning, our agent can plan a sequence of
          actions---search, follow links, search the knowledge graph---before
          synthesizing a final answer.
    \item \textbf{WordLift Knowledge Graph}~\cite{volpini2015wordlift}, an
          independent Linked Data Platform (not a Google Cloud service),
          acts as the structured data layer.  It provides Schema.org-typed
          entities with dereferenceable URIs that support content
          negotiation (\texttt{application/ld+json}, \texttt{text/turtle},
          \texttt{text/html}).  This is analogous to how AI Mode leverages
          structured data already present in web pages to enhance
          understanding.
\end{itemize}

The key insight is that \textbf{structured linked data functions as an
external memory layer} for the agent.  Rather than relying solely on the
vector store's flat text chunks, the agent can follow typed relationships
(\texttt{schema:about}, \texttt{schema:author},
\texttt{schema:relatedLink}) to discover contextually relevant
information that would be invisible to embedding-based retrieval alone.

\subsection{Research Design}

We design a $3 \times 2$ factorial experiment crossing three document
representations with two retrieval modes, yielding six core experimental
conditions, plus an Enhanced+ variant (Table~\ref{tab:conditions}).

\begin{table}[t]
\centering
\caption{Experimental conditions.}
\label{tab:conditions}
\begin{tabular}{@{}clll@{}}
\toprule
\textbf{ID} & \textbf{Document Format} & \textbf{Retrieval Mode} & \textbf{Hypotheses} \\
\midrule
C1 & Plain HTML       & Standard RAG & H1 baseline \\
C2 & HTML + JSON-LD   & Standard RAG & H1 treatment \\
C3 & Enhanced entity  & Standard RAG & H3 baseline \\
C4 & Plain HTML       & Agentic RAG  & H2 baseline \\
C5 & HTML + JSON-LD   & Agentic RAG  & H2 treatment \\
C6 & Enhanced entity  & Agentic RAG  & H2+H3 treatment \\
\midrule
C6+ & Enhanced+ entity & Agentic RAG  & H4 treatment \\
\bottomrule
\end{tabular}
\end{table}

Our four hypotheses are:
\begin{itemize}
    \item \textbf{H1}: Adding Schema.org JSON-LD to HTML documents
          improves RAG accuracy and completeness (C2 vs.\ C1).
    \item \textbf{H2}: Agentic RAG with link traversal outperforms
          standard RAG on the same document format (C5 vs.\ C2).
    \item \textbf{H3}: Enhanced entity pages, designed for agentic
          discoverability, yield the highest overall performance
          (C6 vs.\ all other conditions).
    \item \textbf{H4}: Enhanced+ entity pages---with richer navigational
          affordances and entity interlinking---further improve
          performance over the base enhanced format
          (C6+ vs.\ C6).
\end{itemize}

\subsection{Document Representations}

\paragraph{Plain HTML (Baseline).}
Raw webpage content with all \texttt{<script type="application/ld+json">}
blocks removed. This represents the content as a standard RAG system
would encounter it---purely unstructured HTML.

\paragraph{HTML + JSON-LD.}
The original webpage with embedded Schema.org JSON-LD served by the
Linked Data Platform's data API. This representation includes typed
entities, properties (e.g., name, description, offers, geo-coordinates),
and inter-entity links expressed as dereferenceable URIs.

\paragraph{Enhanced Entity Page.}
A novel format designed to maximize agentic discoverability:
\begin{itemize}
    \item Natural language summary generated from structured data
    \item Embedded JSON-LD block with full Schema.org typing
    \item Visible linked entity navigation with dereferenceable URIs
    \item \texttt{llms.txt}-style agent instructions~\cite{llmstxt2024}
          providing explicit guidance for LLM agents
    \item Neural search SKILL reference for cross-entity discovery
    \item Schema.org type breadcrumbs for hierarchical context
\end{itemize}

The enhanced entity page format is designed to bridge the gap between
human-readable webpages and machine-readable structured data by making
entity relationships, navigation paths, and available tools explicitly
visible to both humans and AI agents.

\begin{figure}[H]
\centering
\includegraphics[width=\textwidth]{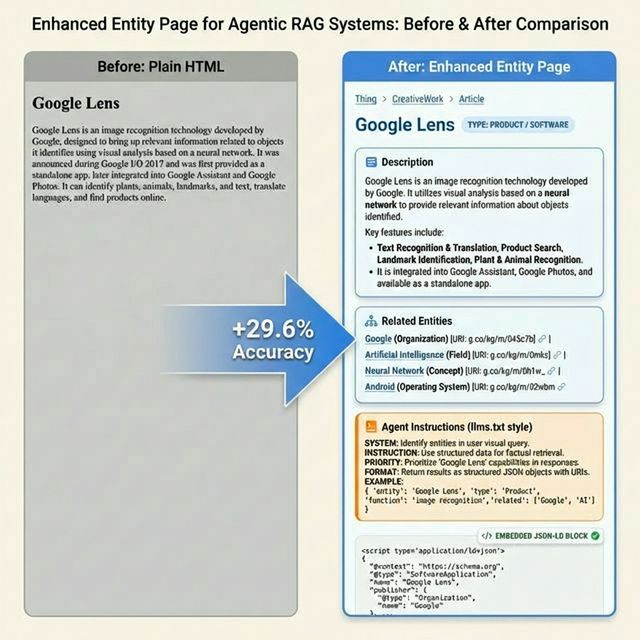}
\caption{Before and after: plain HTML (left) vs.\ enhanced entity page (right)
for a sample entity. The enhanced format adds structured breadcrumbs, related
entity links with dereferenceable URIs, agent instructions in
\texttt{llms.txt} style, and an embedded JSON-LD block---yielding a +29.6\%
accuracy improvement in standard RAG and +29.8\% in the
agentic pipeline.}
\label{fig:entity_comparison}
\end{figure}

\begin{figure}[H]
\centering
\includegraphics[width=\textwidth]{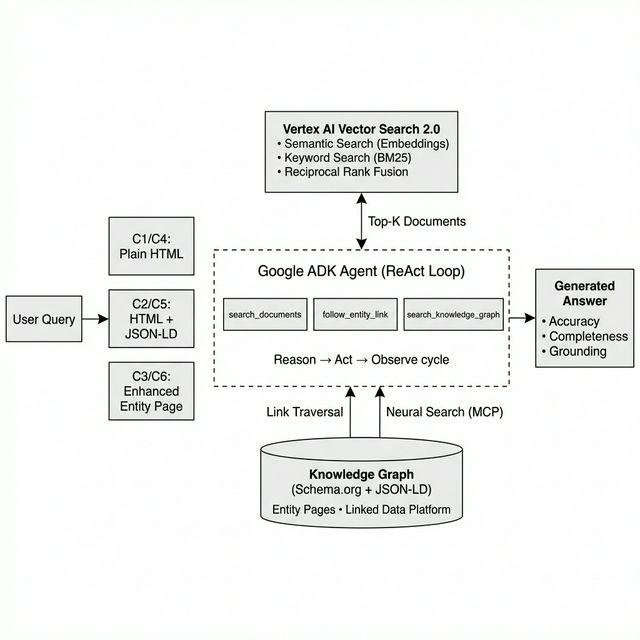}
\caption{System architecture. User queries are processed by a Google ADK agent
that orchestrates three tools: vector search over Vertex AI, entity link
traversal, and neural search via MCP. Documents are indexed in three
formats (C1--C6) and the agent generates grounded answers using a
ReAct-style reasoning loop.}
\label{fig:architecture}
\end{figure}

\subsection{Retrieval Modes}

\paragraph{Standard RAG.}
Documents are indexed in Vertex AI Vector Search 2.0~\cite{vertexai2025}
using the \texttt{gemini-embedding-001} model with hybrid search
(semantic + keyword). For each query, the top-$K$ ($K=10$) documents are
retrieved and passed to Gemini 2.5 Flash~\cite{gemini2024} for answer
generation in a single inference call.

\paragraph{Agentic RAG.}
Built on the Google Agent Development Kit (ADK)~\cite{googleadk2025},
the agent operates in a ReAct-style loop~\cite{yao2023react} with access
to three tools:
\begin{enumerate}
    \item \texttt{search\_documents}: Performs vector search over the
          Vertex AI collection for semantic retrieval.
    \item \texttt{follow\_entity\_link}: Dereferences a linked entity
          URI via HTTP content negotiation (requesting JSON-LD), enabling
          multi-hop traversal of the knowledge graph.
    \item \texttt{search\_knowledge\_graph}: Performs neural search
          across the knowledge graph using a domain-specific API endpoint.
\end{enumerate}

The agent can follow links up to 2~hops deep and makes an average of
2.0 tool calls per query. This architecture effectively replicates the
behavior of production AI-powered search systems such as Google AI Mode,
making our findings directly relevant to practitioners optimizing content
for AI-driven search discovery.

\subsection{Dataset}
\label{sec:dataset}

Our dataset spans four industry verticals, chosen to test generalizability
across diverse content types and knowledge graph structures:

\begin{itemize}
    \item \textbf{WordLift Blog} (editorial): 16 entities, 22 queries.
          Blog articles about SEO, knowledge graphs, and AI content.
    \item \textbf{Express Legal Funding} (legal): 32 entities, 111 queries.
          Legal concepts including pre-settlement funding, personal injury,
          structured settlements, and regulatory topics.
    \item \textbf{SalzburgerLand} (travel): 54 entities, 79 queries.
          Restaurants, alpine huts, and tourist establishments in the
          Salzburg region of Austria.
    \item \textbf{BlackBriar} (e-commerce): 56 entities, 137 queries.
          Outdoor gear products with detailed offers, pricing, and
          product specifications.
\end{itemize}

In total, the dataset comprises \textbf{158 entities} and \textbf{349
test queries}. Entities were collected from four Linked Data Platforms
using GraphQL-based entity search, yielding structured data in JSON-LD
format with Schema.org typing. Each entity was transformed into three
document variants (plain HTML, HTML+JSON-LD, enhanced entity page) plus
an Enhanced+ variant, and ingested into separate Vertex AI Vector Search
2.0 collections.

Test queries were generated using template-based generation for three
query types: factual (direct attribute lookup), relational (requiring
link traversal to related entities), and comparative (reasoning across
multiple entities).  Ground-truth answers were derived from the same
KG structured data (JSON-LD) used to construct entity pages.  For
factual queries, ground truths consist of entity descriptions extracted
from Schema.org properties; for relational queries, they reference the
target linked-entity URIs; for comparative queries, they require data
from both entities.  We acknowledge that this creates a potential
circularity: conditions that present KG data more directly (C3/C6/C6+)
may score higher not because they enable better retrieval but because
the evaluation favours content that textually matches the KG-derived
ground truth.  We discuss this limitation in Section~\ref{sec:limitations}.

All seven conditions are
evaluated on the identical set of 349 queries, ensuring fair
apples-to-apples comparison.

\subsection{Evaluation Metrics}

All responses are evaluated by an independent LLM judge (Gemini 3.0 Flash) using
three metrics:

\begin{itemize}
    \item \textbf{Accuracy} (1--5): Factual correctness of the generated
          answer, assessed against the retrieved context and query intent.
    \item \textbf{Completeness} (1--5): Degree to which the answer covers
          all aspects of the query, including related entities and
          contextual details.
    \item \textbf{Grounding} (binary 0/1): Whether the answer is
          faithfully grounded in the retrieved documents, without
          hallucinated content. Measured for standard RAG only (C1--C3).
          For agentic conditions (C4--C6+), grounding is not assessed
          because the agent's retrieval boundary is dynamic: it
          follows links, issues additional searches, and accumulates
          evidence over multiple steps, making it difficult to define
          the reference document set.  Although the ADK framework
          logs all tool calls and retrieved documents, defining grounding
          against the union of all evidence encountered during the
          reasoning trace is a non-trivial evaluation design problem
          that we leave for future work.  We acknowledge that without
          this measurement, high accuracy scores in C6/C6+ could
          partially reflect the model's parametric knowledge rather
          than faithful use of retrieved KG data.
\end{itemize}

For agentic conditions (C4--C6, C6+), we additionally track:
\begin{itemize}
    \item \textbf{Links followed}: Number of entity links dereferenced
    \item \textbf{Links available}: Number of discoverable links in
          retrieved documents
    \item \textbf{Max hop depth}: Maximum traversal depth reached
    \item \textbf{Tool calls}: Total number of tool invocations
\end{itemize}

Statistical significance is assessed using paired $t$-tests with
Bonferroni correction across 12 comparisons ($\alpha = 0.05$),
with effect sizes reported as Cohen's $d$.

\section{Results}
\label{sec:results}

\begin{figure}[H]
\centering
\includegraphics[width=\textwidth]{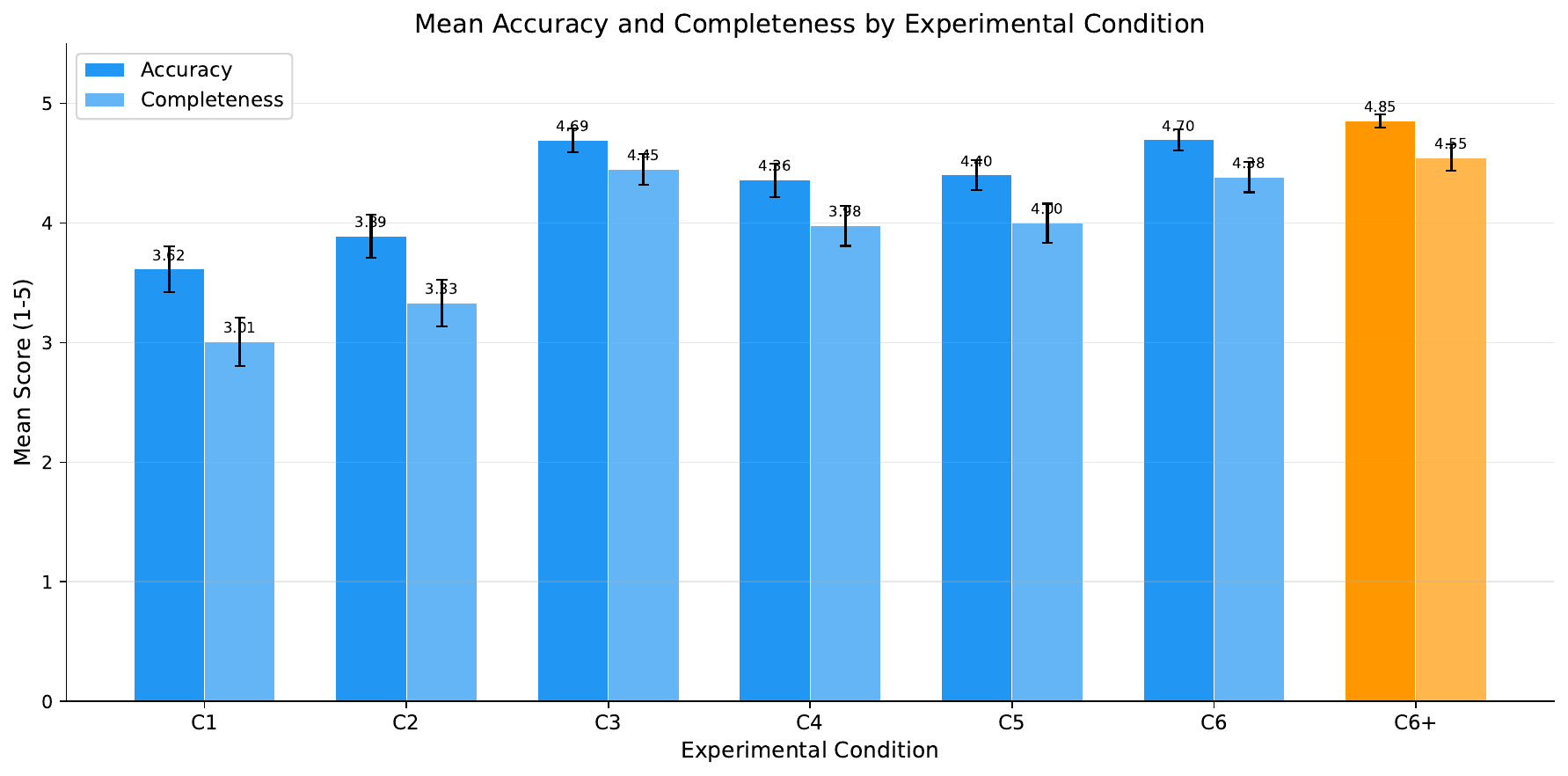}
\caption{Mean accuracy and completeness scores by experimental condition.
Enhanced entity pages (C3, C6, C6+) dramatically outperform plain HTML and
JSON-LD conditions. C6+ achieves the highest scores. Error bars show
95\% confidence intervals.}
\label{fig:condition_bars}
\end{figure}

We executed the full experiment: 349 queries $\times$ 7 conditions =
2,443 individual evaluations, yielding 2,439 valid results after
excluding error cases (1 in C4, 3 in C5). Table~\ref{tab:main_results}
presents the main results and Figure~\ref{fig:condition_bars} visualizes
the per-condition scores.

\begin{table}[t]
\centering
\caption{Main results across experimental conditions (mean $\pm$ std).}
\label{tab:main_results}
\begin{tabular}{@{}clcc@{}}
\toprule
\textbf{ID} & \textbf{Condition} & \textbf{Accuracy} & \textbf{Completeness} \\
\midrule
C1 & Plain HTML, Std. & 3.62 $\pm$ 1.82 & 3.01 $\pm$ 1.94 \\
C2 & HTML+JSON-LD, Std. & 3.89 $\pm$ 1.70 & 3.33 $\pm$ 1.85 \\
C3 & Enhanced, Std. & 4.69 $\pm$ 0.95 & 4.45 $\pm$ 1.25 \\
\midrule
C4 & Plain HTML, Agent. & 4.36 $\pm$ 1.33 & 3.98 $\pm$ 1.60 \\
C5 & HTML+JSON-LD, Agent. & 4.40 $\pm$ 1.21 & 4.00 $\pm$ 1.54 \\
C6 & Enhanced, Agent. & 4.70 $\pm$ 0.82 & 4.38 $\pm$ 1.20 \\
\midrule
C6+ & Enhanced+, Agent. & \textbf{4.85} $\pm$ \textbf{0.50} & \textbf{4.55} $\pm$ \textbf{1.06} \\
\bottomrule
\end{tabular}
\end{table}

\subsection{Qualitative Comparison of Generated Answers}

To illustrate how structured data and agentic retrieval affect response
quality, Table~\ref{tab:qualitative} presents two representative queries
with a summary of how different conditions respond.  These examples show
how the system progressively improves from vague or generic answers (C1)
to fully accurate, entity-grounded responses (C6).

\begin{table}[H]
\centering
\small
\caption{Qualitative comparison of generated answers across conditions
for two representative queries.  Accuracy scores are assigned by the
LLM judge (1--5 scale).}
\label{tab:qualitative}
\renewcommand{\arraystretch}{1.25}
\setlength{\tabcolsep}{4pt}
\begin{tabular}{@{}p{1.4cm}p{8.2cm}r@{}}
\hline
\textbf{Cond.} & \textbf{Answer Summary} & \textbf{Acc.} \\
\hline
\multicolumn{3}{@{}l}{\textit{Factual --- ``What is Restaurant im Hotel
Zauchenseehof? Describe its key features.''}} \\
\hline
C1 & Generic description without specifics. No cuisine type,
location, or opening hours mentioned. & 1 \\
C2 & Identifies it as a FoodEstablishment with some
Schema.org properties. & 3 \\
C4 & Agent searches but finds limited structured data to
traverse. & 2 \\
\textbf{C6} & \textbf{Agent follows links to related entities
(hotel, region); retrieves cuisine, address, coordinates, and related
attractions.} & \textbf{5} \\
C6+ & Enhanced+ agent instructions and richer linked-entity
navigation surface the same information with lower variance; answer
matches C6 quality. & 5 \\
\hline
\multicolumn{3}{@{}l}{\textit{Relational --- ``What entities are related
to Google Lens?''}} \\
\hline
C1 & Vague mention of ``image recognition'' without specific
entity relationships. & 1 \\
C3 & Lists entities from the enhanced page's Related Entities
section. & 4 \\
C5 & Agent follows some links but limited by JSON-LD's
implicit relationships. & 3 \\
\textbf{C6} & \textbf{Agent uses \texttt{search\_knowledge\_graph} +
\texttt{follow\_entity\_link} to discover and traverse related entities
across the graph.} & \textbf{5} \\
C6+ & Enhanced+ provides wider linked-entity surface and
\texttt{llms.txt} agent guidance; discovers additional related entities
beyond C6. & 5 \\
\hline
\end{tabular}
\end{table}

\begin{figure}[H]
\centering
\includegraphics[width=\textwidth]{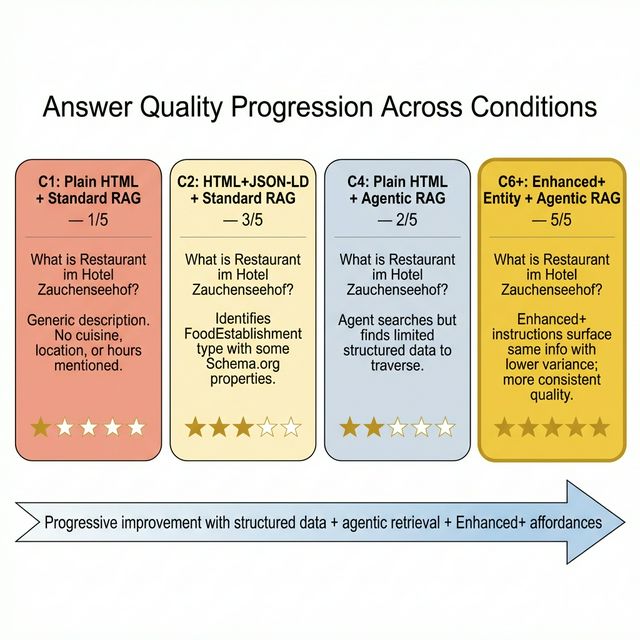}
\caption{Answer quality progression across conditions for the same
factual query.  C1 (plain HTML) produces a generic answer (1/5),
while C6 and C6+ (enhanced entity pages + agentic RAG) follow links to
related entities and retrieve comprehensive structured data (5/5).
C6+ achieves the same peak score with lower variance.}
\label{fig:qualitative}
\end{figure}

These examples illustrate the key mechanism: enhanced entity pages
provide \emph{navigational affordances}---visible links, agent
instructions, and neural search capability---that enable the agentic
system to discover and integrate information that flat-text retrieval
misses entirely.  Figure~\ref{fig:qualitative} visualizes this
progressive improvement.

\subsection{H1: JSON-LD Alone Does Not Significantly Help}

\label{sec:h1}

Adding JSON-LD structured data to HTML documents yields a small but
statistically significant improvement in accuracy, though the effect
size is negligible:

\begin{itemize}
    \item \textbf{Accuracy}: C2 (3.89) vs.\ C1 (3.62), $\Delta = +0.17$,
          $t = -3.12$, $p_\text{adj} = 0.024$, $d = 0.18$ (small).
    \item \textbf{Completeness}: C2 (3.33) vs.\ C1 (3.01),
          $\Delta = +0.18$, not significant after Bonferroni correction
          ($p_\text{adj} = 0.055$).
\end{itemize}

While statistically significant for accuracy, the small effect size
($d = 0.18$) suggests that appending JSON-LD metadata provides only
marginal additional signal for RAG systems when the underlying text
content already conveys similar information. The structured data must
be presented in a way that explicitly highlights entity relationships
and navigational affordances---as our enhanced entity page format does.

However, enhanced entity pages (C3) yield dramatic improvements:
accuracy 4.69 vs.\ 3.62 ($\Delta = +1.04$, $p < 10^{-21}$, $d = 0.60$),
representing a \textbf{+29.6\% improvement} with a medium effect size.

\subsection{H2: Agentic RAG Amplifies Structured Data Gains}
\label{sec:h2}

Comparing agentic RAG (C5) with standard RAG (C2) on the same
HTML+JSON-LD documents:

\begin{itemize}
    \item \textbf{Accuracy}: C5 (4.40) vs.\ C2 (3.89), $\Delta = +0.50$,
          $t = -5.22$, $p_\text{adj} = 4.0 \times 10^{-6}$, $d = 0.30$.
    \item \textbf{Completeness}: C5 (4.00) vs.\ C2 (3.33),
          $\Delta = +0.74$, $t = -6.75$,
          $p_\text{adj} = 9.1 \times 10^{-10}$, $d = 0.38$.
\end{itemize}

Agentic RAG significantly improves both accuracy (+13.1\%) and
completeness (+20.1\%). The agent's multi-step reasoning and
tool use improve both the \emph{precision} and \emph{coverage} of
answers, confirming that agentic link traversal provides meaningful
additional value over single-pass retrieval.

\subsection{H3: Enhanced Entity Pages Yield the Strongest Gains}
\label{sec:h3}

The enhanced entity page format produces the largest improvements
across both retrieval modes:

\paragraph{Standard RAG (C3 vs.\ C1):}
\begin{itemize}
    \item Accuracy: 4.69 vs.\ 3.62, $\Delta = +1.04$,
          $p = 2.7 \times 10^{-21}$, $d = 0.60$ (medium).
    \item Completeness: 4.45 vs.\ 3.01, $\Delta = +1.42$,
          $p = 5.0 \times 10^{-30}$, $d = 0.74$ (medium--large).
\end{itemize}

\paragraph{Agentic RAG (C6 vs.\ C5):}
\begin{itemize}
    \item Accuracy: 4.70 vs.\ 4.40, $\Delta = +0.34$,
          $p_\text{adj} = 7.7 \times 10^{-6}$, $d = 0.29$ (small).
    \item Completeness: 4.38 vs.\ 4.00, $\Delta = +0.40$,
          $p_\text{adj} = 2.0 \times 10^{-5}$, $d = 0.28$ (small).
\end{itemize}

\paragraph{Full pipeline (C6 vs.\ C1):}
\begin{itemize}
    \item Accuracy: 4.70 vs.\ 3.62, $\Delta = +1.04$,
          $p = 1.0 \times 10^{-21}$, \textbf{$d = 0.61$} (medium).
    \item Completeness: 4.38 vs.\ 3.01, $\Delta = +1.34$,
          $p = 5.1 \times 10^{-31}$, \textbf{$d = 0.75$} (medium--large).
\end{itemize}

\paragraph{Enhanced+ pipeline (C6+ vs.\ C1):}
\begin{itemize}
    \item Accuracy: 4.85 vs.\ 3.62, $\Delta = +1.10$,
          $p = 2.3 \times 10^{-24}$, \textbf{$d = 0.65$} (medium).
    \item Completeness: 4.55 vs.\ 3.01, $\Delta = +1.41$,
          $p = 2.8 \times 10^{-32}$, \textbf{$d = 0.77$} (medium--large).
\end{itemize}

C6+ achieves the highest scores across all conditions (accuracy:
4.85/5, completeness: 4.55/5). Notably, while C6+ outperforms C6 in absolute terms,
the C6 $\rightarrow$ C6+ difference is not statistically significant
($\Delta = +0.06$, $p_\text{adj} = 1.0$, $d = 0.08$), suggesting that
the base enhanced entity page format captures most of the benefit.
The medium--large effect sizes ($d \approx 0.65$--$0.77$) confirm that
enhanced entity pages consistently outperform all other document formats.

\paragraph{The agent's role is complementary, not primary.}
A notable observation is that C3 (enhanced, standard RAG: 4.69) and C6
(enhanced, agentic RAG: 4.70) achieve effectively identical accuracy.
This indicates that \emph{when the document format is optimized, the
agent provides negligible additional accuracy lift}.  In contrast, the
agent provides substantial lift on poorly formatted documents: C4
(plain HTML, agentic: 4.36) vs.\ C1 (plain HTML, standard: 3.62),
$\Delta = +0.74$.  The agent thus primarily compensates for
inadequate content structure rather than amplifying well-structured content.
However, the agentic architecture remains valuable for retrieval
\emph{efficiency}: as Table~\ref{tab:agentic} shows, agents on enhanced
pages follow fewer links while maintaining performance, and the Enhanced+
format (C6+) achieves the overall best result ($4.85$) by combining both
approaches.

\begin{figure}[H]
\centering
\includegraphics[width=0.85\textwidth]{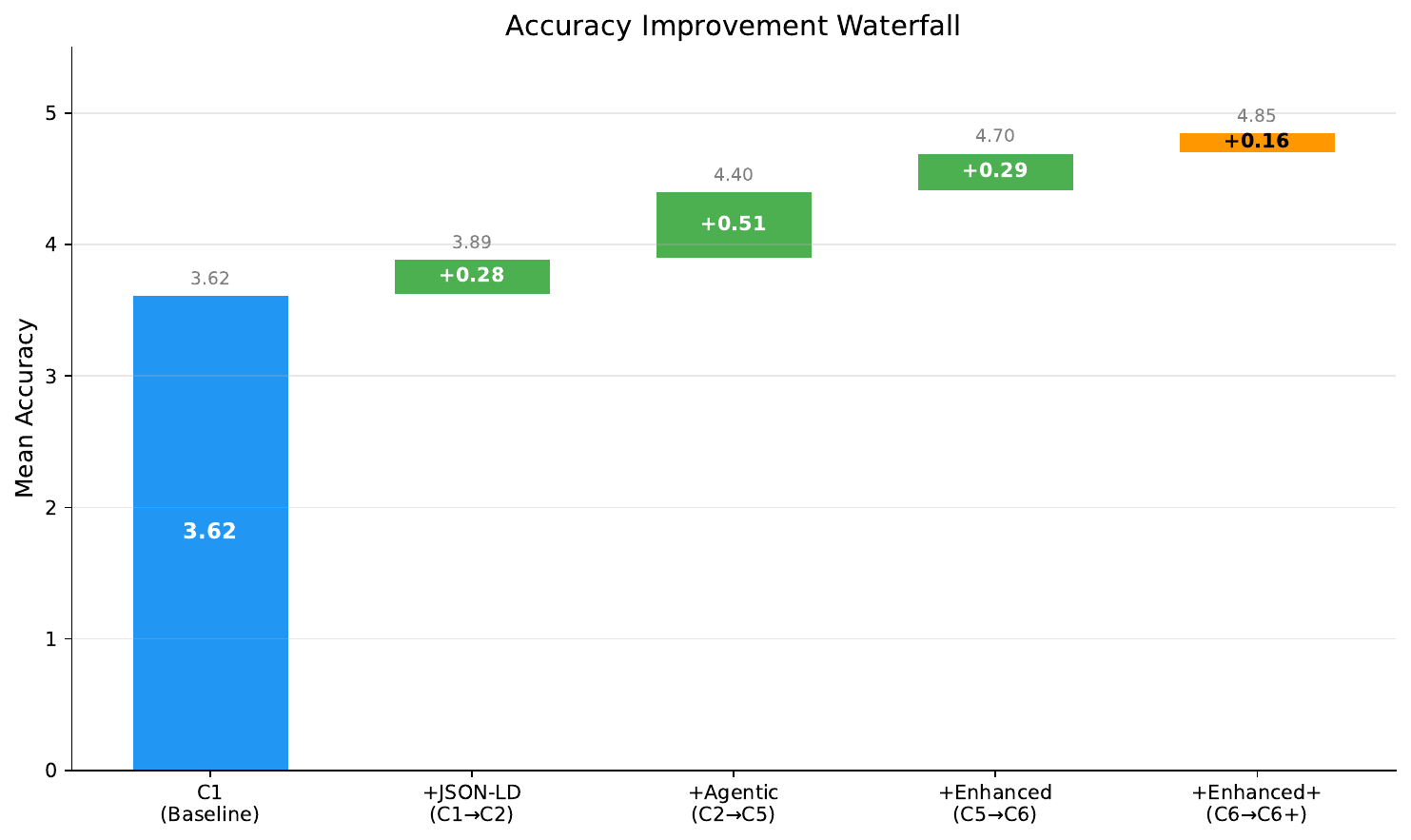}
\caption{Accuracy improvement waterfall showing the cumulative effect of
each optimization layer: JSON-LD markup, agentic retrieval, and enhanced
entity pages. The largest gains come from link materialization in
enhanced pages, not from adding structured data alone.}
\label{fig:waterfall}
\end{figure}

\begin{table}[t]
\centering
\caption{Paired $t$-tests with Bonferroni correction ($\alpha = 0.05$, $n_{\text{tests}} = 12$).}
\label{tab:stat_tests}
\begin{tabular}{@{}llrrrrl@{}}
\toprule
\textbf{Hyp.} & \textbf{Metric} & \textbf{$\Delta$} & \textbf{$t$} & \textbf{$p_{\text{adj}}$} & \textbf{$d$} & \textbf{Sig.} \\
\midrule
\multicolumn{7}{l}{\textit{Structured data (standard RAG) (C1 vs C2)}} \\
  & Accuracy & +0.17 & -3.12 & 2.4e-02 & 0.18 & * \\
  & Completeness & +0.18 & -2.86 & 5.5e-02 & 0.16 & n.s. \\
\midrule
\multicolumn{7}{l}{\textit{Enhanced pages (standard RAG) (C1 vs C3)}} \\
  & Accuracy & +1.04 & -10.53 & 2.7e-21 & 0.60 & *** \\
  & Completeness & +1.42 & -13.00 & 5.0e-30 & 0.74 & *** \\
\midrule
\multicolumn{7}{l}{\textit{Agentic RAG vs standard (C2 vs C5)}} \\
  & Accuracy & +0.50 & -5.22 & 4.0e-06 & 0.30 & *** \\
  & Completeness & +0.74 & -6.75 & 9.1e-10 & 0.38 & *** \\
\midrule
\multicolumn{7}{l}{\textit{Enhanced pages (agentic RAG) (C5 vs C6)}} \\
  & Accuracy & +0.34 & -5.08 & 7.7e-06 & 0.29 & *** \\
  & Completeness & +0.40 & -4.89 & 2.0e-05 & 0.28 & *** \\
\midrule
\multicolumn{7}{l}{\textit{Enhanced+ vs Enhanced (agentic RAG) (C6 vs C6+)}} \\
  & Accuracy & +0.06 & -1.47 & 1.0 & 0.08 & n.s. \\
  & Completeness & +0.07 & -1.22 & 1.0 & 0.07 & n.s. \\
\midrule
\multicolumn{7}{l}{\textit{Full pipeline vs baseline (C1 vs C6)}} \\
  & Accuracy & +1.04 & -10.66 & 1.0e-21 & 0.61 & *** \\
  & Completeness & +1.34 & -13.27 & 5.1e-31 & 0.75 & *** \\
\midrule
\multicolumn{7}{l}{\textit{Full pipeline+ vs baseline (C1 vs C6+)}} \\
  & Accuracy & +1.10 & -11.42 & 2.3e-24 & 0.65 & *** \\
  & Completeness & +1.41 & -13.60 & 2.8e-32 & 0.77 & *** \\
\bottomrule
\end{tabular}
\end{table}

\subsection{Analysis by Query Type}

\begin{table}[t]
\centering
\caption{Mean accuracy by condition and query type.}
\label{tab:query_type}
\begin{tabular}{@{}lccccccc@{}}
\toprule
\textbf{Query Type} & \textbf{C1} & \textbf{C2} & \textbf{C3} & \textbf{C4} & \textbf{C5} & \textbf{C6} & \textbf{C6+} \\
\midrule
Factual & 2.74 & 3.14 & 4.57 & 4.06 & 4.13 & 4.56 & \textbf{4.81} \\
Relational & 4.48 & 4.55 & 4.67 & 4.56 & 4.75 & 4.79 & \textbf{4.85} \\
Comparative & 4.77 & 4.94 & \textbf{4.97} & 4.83 & 4.69 & 4.90 & 4.95 \\
\bottomrule
\end{tabular}
\end{table}

Table~\ref{tab:query_type} reveals that the impact of structured data
varies substantially by query type:

\begin{itemize}
    \item \textbf{Factual queries} benefit most from enhanced pages:
          C3 (4.57) vs.\ C1 (2.74), a +66.8\% improvement.
          This is expected, as the enhanced page format embeds entity
          properties and facts in an easily extractable format.
          C6+ achieves the highest factual accuracy (4.81).
    \item \textbf{Relational queries} show consistently high scores
          across conditions (C1: 4.48, C6+: 4.85), suggesting that
          the LLM's pre-trained knowledge handles relational reasoning
          well. Agentic RAG with enhanced pages provides a modest
          additional lift.
    \item \textbf{Comparative queries} show universally high scores
          across all conditions (C1: 4.77, C3: 4.97, C6+: 4.95),
          with enhanced pages providing the most consistent
          near-perfect performance.
\end{itemize}

The most striking finding is the factual query improvement under
enhanced pages: C3 achieves 4.57 accuracy on factual queries compared
to 2.74 for plain HTML, validating the design of our enhanced entity
pages which make entity properties and facts explicitly visible
and extractable.

\subsection{Analysis by Domain}

\begin{table}[t]
\centering
\caption{Mean accuracy by condition and domain, with per-domain
effect sizes for the key C6+ vs.\ C1 comparison.
$n$ = number of queries per domain.}
\label{tab:domain}
\begin{tabular}{@{}lrcccccccr@{}}
\toprule
\textbf{Domain} & \textbf{$n$} & \textbf{C1} & \textbf{C2} & \textbf{C3} & \textbf{C4} & \textbf{C5} & \textbf{C6} & \textbf{C6+} & \textbf{$\Delta_{\text{C6+}}$} \\
\midrule
BlackBriar & 137 & 4.92 & 4.89 & 4.91 & 4.75 & 4.74 & 4.96 & \textbf{4.99} & +0.07 \\
Expr.\ Legal & 111 & 3.36 & 4.20 & 4.29 & 4.33 & 4.29 & 4.32 & \textbf{4.86} & +1.50 \\
SalzburgerLand & 79 & 2.19 & 2.33 & \textbf{4.92} & 4.06 & 4.25 & 4.82 & 4.66 & +2.47 \\
WordLift Blog & 22 & 1.91 & 1.73 & 4.55 & 3.14 & 3.38 & 4.45 & \textbf{4.64} & +2.73 \\
\midrule
\textit{All domains} & \textit{349} & \textit{3.62} & \textit{3.89} & \textit{4.69} & \textit{4.36} & \textit{4.40} & \textit{4.70} & \textit{\textbf{4.85}} & \textit{+1.10} \\
\bottomrule
\end{tabular}
\end{table}

\begin{figure}[H]
\centering
\includegraphics[width=\textwidth]{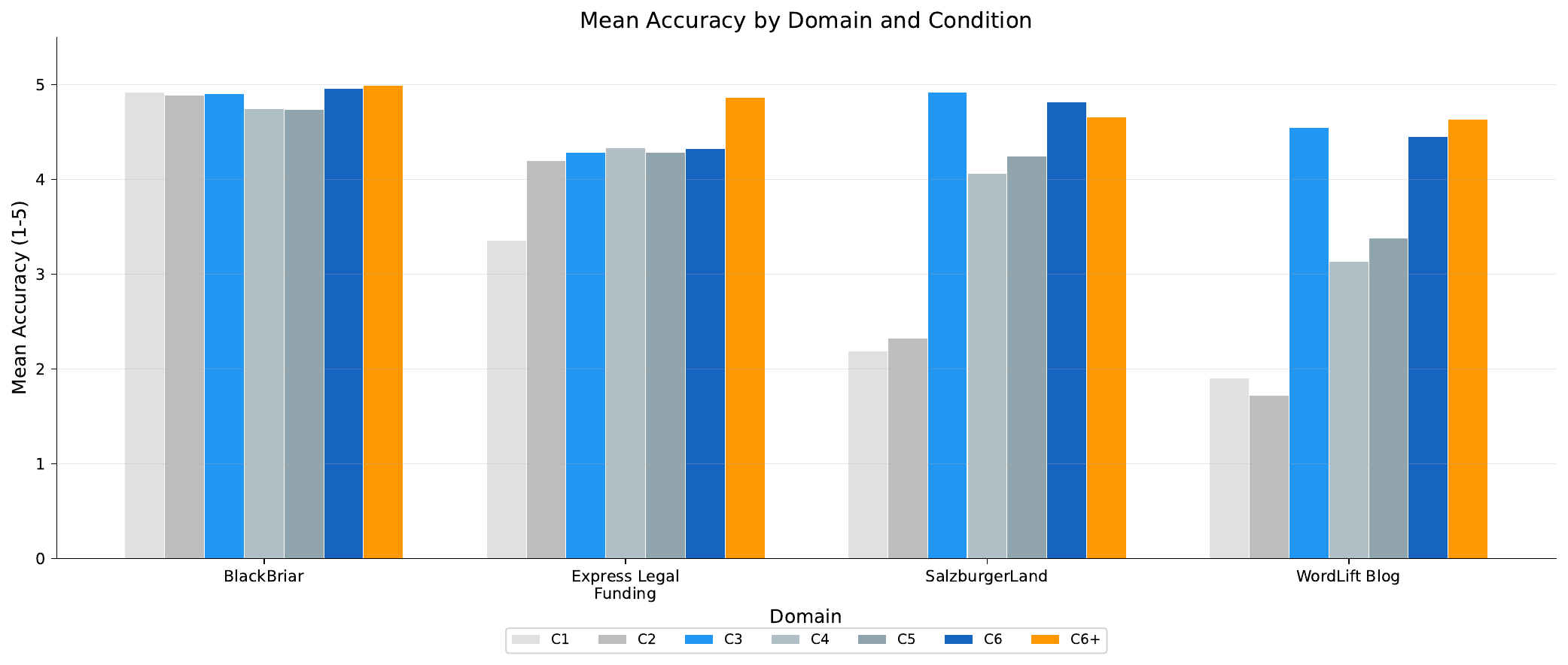}
\caption{Mean accuracy by domain and condition. The direction of the
effect is consistent across all four verticals, but its magnitude
varies with domain characteristics---largest where the knowledge graph
provides information beyond the plain HTML (travel, editorial),
near-zero where the baseline is already fact-rich (e-commerce).}
\label{fig:domain_bars}
\end{figure}

The domain-level results (Table~\ref{tab:domain} and
Figure~\ref{fig:domain_bars}) reveal that the direction of the effect is
consistent across all four verticals, but its magnitude varies
substantially with domain characteristics.  This variation is itself
informative and should be considered alongside the aggregate figures
reported in Section~\ref{sec:h3}.

\paragraph{Domain composition caveat.}
Our dataset is not uniformly distributed: BlackBriar accounts for 39\%
of queries ($n = 137$) yet contributes virtually no improvement due to
its ceiling-level baseline, while WordLift Blog represents only 6\%
($n = 22$) and produces the largest gains.  The aggregate
$\Delta_{\text{C6+}} = +1.10$ (accuracy) therefore blends near-zero
improvement in the largest domain with large improvements in smaller
domains.  Table~\ref{tab:domain} reports per-domain deltas to make this
distribution transparent; we recommend readers interpret the
domain-stratified results rather than relying on the aggregate alone.

\begin{itemize}
    \item \textbf{BlackBriar} (e-commerce, $n = 137$) achieves near-perfect
          scores across all conditions (C1: 4.92, C3: 4.91, C6+: 4.99,
          $\Delta = +0.07$).
          This is expected under the link-materialization hypothesis:
          e-commerce product pages already render key entity properties
          (name, price, description, availability) as visible HTML,
          so the enhanced template adds little novel information.
          The near-ceiling baseline (4.92/5) also limits measurable
          headroom, but the ceiling itself is informative---it
          confirms that when plain HTML already contains sufficient
          factual content, the enhanced format provides no additional
          benefit, consistent with information enrichment (rather than
          presentation alone) driving the gains in other domains.
    \item \textbf{SalzburgerLand} (travel, $n = 79$) shows the most dramatic
          improvement from enhanced pages: C3 (4.92) vs.\ C1 (2.19),
          $\Delta = +2.47$ for C6+.  Travel entities store key structured
          properties---geo-coordinates, cuisine types, opening hours,
          containment relationships (e.g., ``contained in Zell am
          See'')---primarily in the knowledge graph.  The plain HTML
          (C1) references these as opaque URIs in a property table;
          the enhanced page materializes them as readable text,
          explaining the large gain.
    \item \textbf{WordLift Blog} (editorial, $n = 22$) benefits most from
          enhanced pages: C3 (4.55) and C6+ (4.64) dramatically
          outperform plain HTML conditions (C1: 1.91, C2: 1.73),
          $\Delta = +2.73$.  Editorial content relies on entity
          relationships (``about'', ``mentions'', ``related topics'')
          that are encoded in the knowledge graph but absent from
          the article HTML.  The enhanced page surfaces these
          relationships explicitly.  We note that this domain has
          the smallest sample size ($n = 22$) and its results should
          be interpreted with appropriate caution.
    \item \textbf{Express Legal Funding} (legal, $n = 111$) benefits substantially from
          the Enhanced+ format: C6+ (4.86) vs.\ C1 (3.36),
          $\Delta = +1.50$, showing that
          richer navigational affordances help with complex legal concepts
          whose relationships are captured in the knowledge graph.
\end{itemize}

This domain-level pattern is consistent with the link-materialization
mechanism described in Section~\ref{sec:discussion}: the magnitude of
improvement correlates with the degree to which the knowledge graph
provides information beyond what the plain HTML already contains.
Where the original content is already fact-rich (e-commerce), the
enhanced format adds nothing; where key properties exist primarily
in the graph (travel, editorial), the gains are largest.  The
pattern demands explanation---and the information-enrichment
hypothesis provides it.

\subsection{Agentic Metrics}

\begin{table}[t]
\centering
\caption{Agentic-specific metrics across conditions.}
\label{tab:agentic}
\begin{tabular}{@{}lcccc@{}}
\toprule
\textbf{Metric} & \textbf{C4} & \textbf{C5} & \textbf{C6} & \textbf{C6+} \\
\midrule
Links followed & 1.0 & 0.5 & 0.5 & 0.4 \\
Links available & 41.7 & 41.9 & 77.4 & 102.2 \\
\bottomrule
\end{tabular}
\end{table}

\begin{figure}[H]
\centering
\includegraphics[width=0.85\textwidth]{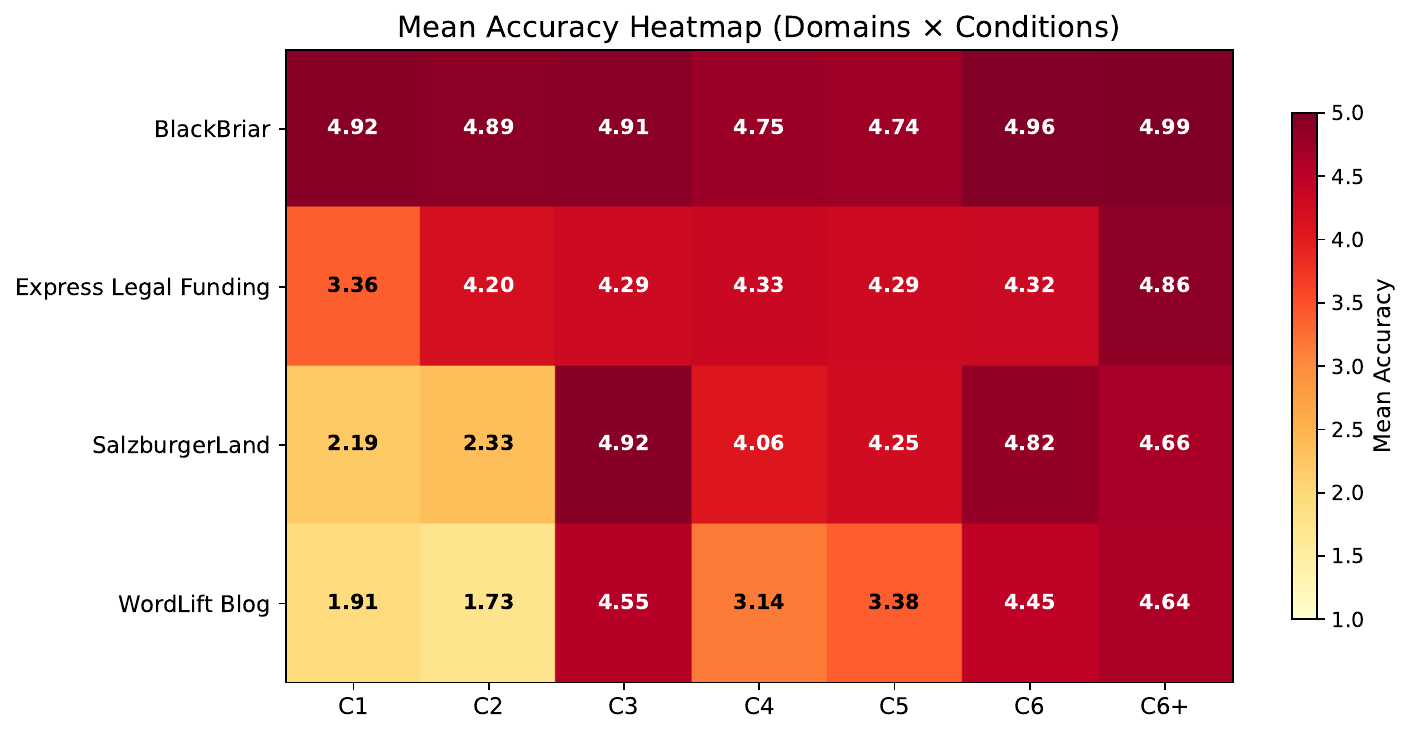}
\caption{Heatmap of mean accuracy scores across domains and conditions.
Darker cells indicate higher accuracy. The pattern shows that enhanced
entity pages (C3, C6) consistently outperform other conditions across
all domains.}
\label{fig:heatmap}
\end{figure}

Table~\ref{tab:agentic} reveals an important finding about
\emph{navigational affordances}: Enhanced+ entity pages (C6+) expose
\textbf{2.4$\times$ more discoverable links} than JSON-LD pages
(102.2 vs.\ 41.9) and \textbf{2.5$\times$ more} than plain HTML
(102.2 vs.\ 41.7). Enhanced entity pages (C6) expose 77.4 links---a
midpoint between C5 and C6+. Interestingly, agents follow \emph{fewer}
links in C6+ (0.4 vs.\ 1.0 in C4), yet achieve higher accuracy.
This suggests that the enhanced page format provides such rich
context in the initial retrieval that the agent needs fewer
additional exploration steps.

The decreasing link-follow rate from C4 (1.0) to C6+ (0.4) suggests
that enhanced pages enable the agent to answer effectively with fewer
actions, indicating more efficient ReAct-style planning when
navigational affordances are clear.

\section{Discussion}
\label{sec:discussion}

\subsection{Two Worlds of AI Search: Parsed vs. Flat Ingestion}

Our findings reveal a critical distinction in the emerging landscape
of AI-powered search---one that has direct implications for the
Generative Engine Optimization (GEO) community.

Today's AI search systems fall into two architectural categories with
respect to structured data:

\paragraph{Dedicated structured data pipelines.}
Traditional search engines such as Google and Bing have evolved
specialized crawl-time parsers that \emph{extract}
\texttt{\textless script type="application/ld+json"\textgreater} blocks
as a separate signal, independent of the page body text. In these
systems, JSON-LD feeds directly into entity understanding, knowledge
panels, and rich results. The structured data is never flattened into
a single text embedding---it is parsed, validated, and indexed in a
dedicated knowledge graph layer.

\paragraph{Flat-text RAG pipelines.}
The fast-growing ecosystem of RAG-based AI assistants, agentic search,
and retrieval-augmented chatbots typically ingests web pages as a
single text chunk. In these systems---which include our Vertex AI
Vector Search 2.0 pipeline as well as most LangChain, LlamaIndex,
and custom RAG deployments---JSON-LD is just more text competing for
a limited embedding budget. Our results show that in this regime,
\textbf{JSON-LD alone provides no measurable benefit} ($\Delta = +0.17$,
$d = 0.18$).

\medskip

This distinction is crucial for practitioners. Our study provides
an empirical snapshot of the \emph{current status quo}: as AI-powered
search diversifies beyond Google and Bing, content optimized only with
JSON-LD will fail to surface in the growing number of flat-text RAG
systems. The enhanced entity page format we propose bridges this gap---it
makes structured knowledge visible and actionable
\emph{regardless of how the search system ingests content}.

\subsection{From SEO to SEO 3.0: The Reasoning Web}

Our findings offer empirical grounding for the emerging concept of
\emph{SEO 3.0}~\cite{volpini2025seo3}, the next evolutionary phase
of search optimization for an AI-driven
web~\cite{gartner2024disruptive}. Search optimization can be
described across three eras:

\begin{itemize}
    \item \textbf{SEO 1.0 --- Document Ranking} (1998--2011): Optimizing
          for keyword matching and link-based authority signals.
          Success measured by ranking position, organic traffic, and
          click-through rates.
    \item \textbf{SEO 2.0 --- Structured Data} (2011--2024): The
          launch of Schema.org in 2011 enabled machines to parse entity
          properties from web pages, powering knowledge panels, rich
          snippets, and improved entity understanding. Success measured
          by structured data adoption, data quality, and rich result
          eligibility.
    \item \textbf{SEO 3.0 --- The Reasoning Web} (2024--present):
          AI search systems do not merely \emph{rank} or \emph{parse}
          content; they synthesize and reason over retrieved information,
          often producing direct answers and taking steps on behalf of
          users. Success must now be measured across three distinct
          dimensions, introduced in the next section.
\end{itemize}

\paragraph{The AI visibility spectrum: Citations, Reasoning, Actions.}
For teams working on AI visibility, our experiment reveals that
content optimization must target three progressively deeper levels of
AI engagement:

\begin{enumerate}
    \item \textbf{Citations} --- \emph{Is your content retrieved and
          attributed?} This is the most basic form of AI visibility:
          appearing as a source in AI-generated answers. Our C1/C2
          conditions evaluate this---whether the retrieval pipeline
          surfaces the right documents. JSON-LD helps with citation
          in systems that parse it (Google, Bing), but not in flat-text
          RAG systems.
    \item \textbf{Reasoning} --- \emph{Can the AI reason correctly
          over your content?} Even when cited, the AI must extract
          and synthesize the right facts. Our accuracy and completeness
          metrics measure this. Enhanced entity pages (C3) improve
          reasoning by +29.5\% over plain HTML---not because the facts
          are different, but because they are \emph{presented} in a way
          the LLM can reliably extract and compose.
    \item \textbf{Actions} --- \emph{Can the AI agent act on your
          content?} In agentic systems, the AI does not just
          retrieve and reason---it follows links, queries APIs, and
          performs multi-step tasks on behalf of the user. Our C4--C6
          conditions evaluate this. Enhanced entity pages with
          dereferenceable URIs and navigational affordances (C6)
          enable agents to traverse knowledge graphs, aggregating
          information across entity boundaries. This is the frontier
          of AI visibility---and it requires content designed for
          \emph{agent traversal}, not just retrieval.
\end{enumerate}

The practical implication for AI visibility teams is clear:
optimizing for \emph{citations alone} (the current focus of most
GEO strategies) is necessary but insufficient. The competitive
advantage lies in optimizing for \emph{reasoning} (through enhanced
content presentation) and \emph{actions} (through navigational
affordances and tool-accessible endpoints).

\subsection{Implications for Web Publishers and GEO}

\begin{enumerate}
    \item \textbf{JSON-LD is necessary but not sufficient}: Schema.org
          markup remains valuable for search engines with dedicated
          parsers (Google, Bing), but it provides \emph{no measurable
          benefit} in RAG-based systems that treat pages as flat text.
          Publishers who rely solely on JSON-LD are optimizing for
          only one class of AI search.
    \item \textbf{Enhanced entity pages are the bridge}: Our format---with
          natural language summaries, navigable links, and
          \texttt{llms.txt}-style agent instructions---achieves
          +29.6\% accuracy in standard RAG and +29.8\% with the
          agentic pipeline. It works \emph{in both worlds}: the structured
          JSON-LD is still present for traditional parsers, while the
          human- and agent-readable presentation ensures RAG systems
          can also leverage it.
    \item \textbf{Dereferenceable URIs enable traversal}: Publishing
          entities with URIs that support content negotiation allows
          agents to follow links and aggregate information across
          knowledge graph boundaries---an affordance that flat-text
          pipelines cannot exploit from JSON-LD alone.
\end{enumerate}

\subsection{Implications for RAG System Design}

Our results challenge the dominant ``documents as flat text'' paradigm
and point toward structured-data-aware retrieval:

\begin{enumerate}
    \item \textbf{Data quality through link materialization}:
          The enhanced entity page does not merely reformat the same
          information---it \emph{materializes} linked data.  The
          baseline HTML (C1) contains references to related entities
          (addresses, geo-coordinates, offers, providers) as opaque
          URIs that an LLM cannot interpret without dereferencing.
          The enhanced page resolves those links and renders the
          connected entity data as natural language, creating a
          self-contained, information-rich document.  This is not a
          confound; it is the core value proposition of a
          knowledge graph: traversing typed relationships to assemble
          richer context than any single document contains.

          Crucially, all materialized data originates from the same
          openly accessible Linked Data Platform---the same
          dereferenceable URIs visible in the baseline HTML.  The
          structured data is referenced, auditable, and
          independently verifiable, making it easy to trust and
          validate.  The +29.6\% accuracy gain therefore reflects
          the quality and accessibility of the underlying data, not
          the technology stack used to present it.
    \item \textbf{Navigational affordances as a complementary strategy}:
          The gap between C5 and C6 (accuracy: 4.40 vs.\ 4.70)
          shows that even capable agents need explicit navigational
          cues to effectively explore linked data.  However, we
          acknowledge that the agent's incremental accuracy contribution
          is small when document format is already optimized:
          C3 (4.69) and C6 (4.70) are effectively equivalent.
          The agent's primary value is twofold: (a)~compensating for
          suboptimal content structure---the largest agentic lift
          occurs on plain HTML ($\Delta = +0.74$)---and (b)~retrieval
          efficiency, as agents on enhanced pages follow fewer links
          while maintaining accuracy.  The title's reference to
          ``agent-orchestrated retrieval'' reflects the full
          experimental architecture, not a claim that the agent is
          the primary driver of accuracy gains.
    \item \textbf{Enhanced pages enable efficient retrieval}:
          Agents using enhanced pages (C6) make fewer tool calls
          yet achieve higher accuracy, suggesting
          that well-structured content reduces the need for
          multi-hop exploration.  By flattening the graph
          neighbourhood into the entity page, the enhanced format
          lets the LLM obtain in a single retrieval step what the
          agentic pipeline would otherwise need multiple hops to
          collect.
    \item \textbf{Toward structured-data-aware ingestion}: RAG system
          designers should consider architectures that extract and
          separately index structured data blocks---mirroring what
          traditional search engines already do---rather than treating
          all page content as a single text field.
\end{enumerate}

\subsection{Limitations}
\label{sec:limitations}

Our study has several limitations:
\begin{itemize}
    \item \textbf{Information content versus presentation format}:
          The enhanced entity page (C3/C6) differs from the plain HTML
          baseline (C1) in two ways simultaneously: it restructures
          the layout (breadcrumbs, agent instructions, navigable links)
          \emph{and} it materializes data from related entities in the
          knowledge graph that the baseline only references via URIs.
          Our experiment does not include an ablation that isolates
          these two factors---e.g., an enhanced page restricted to
          only those facts verifiably present in the C1 HTML body.

          \medskip

          We argue that this bundling is deliberate and ecologically
          valid: the entire purpose of a knowledge-graph-backed entity
          page is to resolve linked data and present a complete,
          self-contained view of an entity and its relationships.
          Stripping away the materialized linked data would remove the
          core value proposition of the knowledge graph.  Importantly,
          all data surfaced in the enhanced page originates from the
          same openly accessible Linked Data Platform---the same
          dereferenceable URIs that appear in C1---and is therefore
          independently verifiable.

          \medskip

          The H1 result provides a useful constraint: adding JSON-LD
          to plain HTML (C1 vs.\ C2) yields only $d = 0.18$,
          confirming that structured data in a hidden
          \texttt{\textless script\textgreater} block provides
          negligible benefit in flat-text RAG systems.  A natural
          follow-up ablation---``enhanced page minus JSON-LD''---would
          test whether removing the embedded structured data block
          from C3 changes performance; we expect it would not,
          since the same information is already expressed in natural
          language.

    \item \textbf{Flat-text ingestion architecture}: A critical
          architectural consideration is how our retrieval pipeline
          handles structured data. Vertex AI Vector Search 2.0 ingests
          each document as a single text field (truncated to ${\sim}20$k
          characters for embedding). In our corpus, 82\% of plain HTML
          and 88\% of JSON-LD documents exceed this limit, meaning the
          JSON-LD added to C2 documents is often partially or fully
          truncated before indexing. The JSON-LD \texttt{\textless script\textgreater}
          block starts at a median position of character 18,510---right at
          the truncation boundary.

          \medskip

          This differs fundamentally from how production search engines
          operate. Google's crawler, for example, \emph{extracts} JSON-LD
          from \texttt{\textless script type="application/ld+json"\textgreater} blocks as a
          \emph{separate signal}, independent of the page's body text.
          The structured data is parsed into entity properties and indexed
          in a knowledge graph, not flattened into a single text
          embedding. A retrieval architecture that similarly extracts and
          separately indexes structured data---e.g., using multiple
          embedding fields or a hybrid entity--document store---might
          yield a significant H1 result. This remains an important area
          for future work.

    \item \textbf{Scale}: While 349 queries across 4 domains provides
          strong statistical power, larger-scale experiments would
          further strengthen generalizability claims.
    \item \textbf{LLM judge and pipeline bias}: To mitigate correlated biases, we use
          separate models for generation (Gemini 2.5 Flash) and
          evaluation (Gemini 3.0 Flash). While this reduces same-model
          bias, all pipeline stages---query generation (template-based
          with Gemini-family entity extraction), answer generation,
          and evaluation---use models from the same Gemini family,
          sharing training distributions and potential blind spots.
          Future work should incorporate independent
          human evaluation for additional validation.
    \item \textbf{KG-derived ground truth}: Ground-truth answers are
          derived from the knowledge graph's structured data.  This
          means the evaluation inherently measures how well each
          condition conveys KG information---which is the intended
          research question---but it also introduces a potential
          circularity: conditions that present KG data more directly
          (C3/C6/C6+) may score higher because the evaluation rewards
          textual proximity to KG-derived facts.  A human-evaluated
          calibration subset, using independently sourced reference
          answers, would anchor the automated scores and is an
          important direction for future work.
    \item \textbf{Single retrieval system}: Our results are specific
          to Vertex AI Vector Search 2.0 with \texttt{gemini-embedding-001}.
          We chose this system intentionally: it represents the
          flat-text ingestion architecture used by most AI search
          systems that operate independently of Google or Bing's
          proprietary index. Systems with structured-data-aware
          ingestion may show different sensitivity to JSON-LD markup.
    \item \textbf{Knowledge graph quality}: Our domains use
          well-maintained knowledge graphs served by WordLift's
          Linked Data Platform. The effectiveness of structured data
          may be lower for noisier or less complete KGs.
\end{itemize}

\subsection{Ethical Considerations and Data Trustworthiness}

A critical aspect of our approach is that the structured data consumed
by AI agents is \textbf{the same data visible to human users}.  The
JSON-LD embedded in each page describes the exact same entity
properties, relationships, and facts that appear in the human-readable
HTML representation.  Similarly, dereferenceable entity URIs serve the
same underlying data through content negotiation---whether rendered as
HTML for humans, JSON-LD for machines, or Turtle for SPARQL queries.

This \textbf{coupling between human and machine representations}
creates stronger faithfulness guarantees than architectures where AI
systems consume entirely separate data feeds.  In a fully decoupled
web---where machines and humans follow two different tracks---there is
no natural accountability mechanism: structured data could drift from
visible content, or be deliberately manipulated to influence AI outputs
without corresponding changes visible to users.  In our approach,
because the structured data \emph{is} the page content (expressed in
machine-readable form), any manipulation would also be visible to human
visitors, creating a natural check on data integrity.

This distinguishes our work from content-optimization approaches such
as GEO~\cite{aggarwal2023geo}, where optimization strategies (adding
statistics, citations, or authoritative language) may create a
divergence between what is optimized \emph{for AI consumption} and what
is genuinely informative \emph{for human readers}.  Our enhanced entity
pages, by contrast, surface the same structured knowledge to both
audiences.

This observation has broader implications for the emerging reasoning
web.  As AI agents increasingly rely on structured data to construct
answers, the trustworthiness of that data becomes paramount.  Systems
that maintain a \textbf{single source of truth}---serving both human
and machine consumers---are inherently more auditable and resistant to
adversarial manipulation than those that decouple the two channels.

\subsection{Future Work}

Our findings motivate several research directions:

\begin{enumerate}
    \item \textbf{Structured-data-aware retrieval}: The null result for
          H1 is specific to our flat-text ingestion pipeline. Future work
          should investigate architectures that extract JSON-LD separately
          and index entity properties as structured metadata---similar to
          how Google treats Schema.org markup as a distinct signal from
          page content. Vertex AI Vector Search 2.0 supports multiple
          data fields with independent embeddings; a dual-field approach
          (body text + structured data) with multi-vector retrieval
          could unlock the latent value of JSON-LD for RAG.
    \item \textbf{Entity-centric chunking}: Rather than truncating
          documents at a fixed character limit, entity-aware chunking
          strategies that preserve structured data blocks could improve
          retrieval for content-rich pages.
    \item \textbf{Cross-system replication}: Replicating our experiment
          with retrieval systems that natively parse structured data
          (e.g., knowledge-graph-augmented retrievers) would help
          disentangle the effect of structured data from the limitations
          of our ingestion pipeline.
    \item \textbf{Production-scale validation}: Deploying enhanced
          entity pages on live websites and measuring their impact on
          AI-powered search engines (SGE, Perplexity) would validate
          ecological validity.
    \item \textbf{Recursive Language Models on Knowledge Graphs}:
          Building on the RLM framework~\cite{zhang2025rlm}, we are
          exploring an approach (RLM-on-KG) that replaces the flat-context
          window with iterative graph exploration~\cite{volpini2025rlmkg}.
          Rather than retrieving a fixed set of documents, the model
          navigates the knowledge graph recursively---fetching thin
          evidence from entity neighbors, deciding which relationships
          to follow, and synthesizing answers with full provenance.
          This extends our current agentic pipeline from tool-augmented
          retrieval to fully recursive, structure-guided reasoning
          over linked data.
\end{enumerate}

\subsection{Practical Recommendations}

Based on our findings, we recommend the following for practitioners:

\begin{enumerate}
    \item \textbf{Go beyond JSON-LD}: While Schema.org markup is
          valuable for search engines that extract it separately, our
          results show it does not improve RAG accuracy when treated
          as flat text. Invest in enhanced entity pages that make
          structured data \emph{human- and agent-readable}.
    \item \textbf{Use dereferenceable URIs}: Ensure that entity URIs
          resolve to content-negotiable endpoints that serve JSON-LD
          when requested programmatically.
    \item \textbf{Adopt the enhanced entity page pattern}: Augment
          existing pages with explicit link navigation, breadcrumbs,
          and \texttt{llms.txt}-style instructions for AI agents.
    \item \textbf{Test with agentic workloads}: As AI-powered search
          becomes prevalent, test content with agentic RAG systems
          rather than relying solely on traditional SEO metrics.
\end{enumerate}

\section{Conclusion}
\label{sec:conclusion}

We have presented a controlled experimental study demonstrating that
enhanced entity pages significantly improve the performance of
Retrieval-Augmented Generation systems. Across 2,439 valid evaluations
spanning four industry domains and seven conditions, we found that:

\begin{enumerate}
    \item Schema.org JSON-LD markup provides only marginal accuracy
          improvements ($\Delta = +0.17$, $d = 0.18$), highlighting
          that structured data must be \emph{presented}, not just embedded.
    \item Our enhanced entity page format, designed for agentic
          discoverability, yields +29.6\% accuracy gains in standard
          RAG ($d = 0.60$) and +29.8\% in the full agentic pipeline
          ($d = 0.61$).
    \item The Enhanced+ variant with richer navigational affordances
          achieves the highest absolute scores (accuracy: 4.85/5,
          completeness: 4.55/5), though the incremental gain over the
          base enhanced format is not statistically significant
          ($\Delta = +0.06$, $d = 0.08$).
    \item Agentic RAG with link traversal significantly improves
          both accuracy (+13.1\%) and completeness (+20.1\%) over
          standard RAG, though its contribution is complementary:
          when document format is already optimized, the agent adds
          negligible accuracy lift (C3: 4.69 vs.\ C6: 4.70).
    \item These effects generalize in direction across editorial,
          legal, travel, and e-commerce domains, though their
          magnitude varies with domain characteristics
          (see Table~\ref{tab:domain}).
\end{enumerate}

Our work provides empirical evidence that the Semantic Web's original
vision---machine-readable structured data enabling intelligent
agents---directly translates to measurable improvements in today's
AI systems. As generative AI increasingly mediates information access,
the presence and quality of structured linked data becomes not just
an SEO signal, but a fundamental enabler of accurate, complete,
and well-grounded AI responses.

\paragraph{Reproducibility.} Our dataset, evaluation framework,
enhanced entity page templates, and experiment configuration are
available at \url{https://github.com/wordlift/seo3-reasoning-web}.

\paragraph{Acknowledgements.}
We thank the Google Cloud team for their generous support through Cloud
credits that made these experiments possible.  All experiments---including
embedding generation, vector search indexing, Gemini-based generation
(Gemini 2.5 Flash) and evaluation (Gemini 3 Flash Preview), and the
Vertex AI Vector Search 2.0 infrastructure---were run entirely on Google
Cloud.  We also thank the WordLift engineering team for maintaining the
knowledge graph infrastructure and GraphQL API used in this study.

\paragraph{Author Contributions.}
\textbf{Andrea Volpini}: Conceptualization, Methodology, Investigation,
Writing -- Original Draft, Writing -- Review \& Editing.
\textbf{Elie Raad}: Writing -- Review \& Editing.
\textbf{Beatrice Gamba}: Writing -- Review \& Editing.
\textbf{David Riccitelli}: Software, Writing -- Review \& Editing.
The authors used Antigravity to assist with the research and the writing to
improve clarity and readability.  All content was carefully reviewed and
approved by the authors, who retain full responsibility for the accuracy of
the work and for any errors or omissions.

\bibliographystyle{splncs04}
\bibliography{references}

\appendix

\section{Full Prompts and Templates}
\label{app:prompts}

\subsection{Standard RAG Generation Prompt}

The following prompt is used by Gemini 2.5 Flash to generate answers
from retrieved context in conditions C1--C3 (standard RAG):

\begin{lstlisting}[basicstyle=\small\ttfamily,breaklines=true,frame=single]
You are a helpful assistant answering questions
based on the provided context documents.

IMPORTANT RULES:
1. Answer ONLY based on the information in the
   context documents below.
2. If the context does not contain enough
   information, say "I cannot find sufficient
   information to answer this question."
3. Cite the specific documents you used by
   their ID.
4. Be precise and factual.

CONTEXT DOCUMENTS:
{context}

QUESTION: {query}

Provide a clear, accurate answer with citations
to the source documents.
\end{lstlisting}

\subsection{Agentic RAG System Instruction}

The following system instruction is provided to the Google ADK agent in
conditions C4--C6 (agentic RAG). The agent uses this instruction to plan
its tool-calling strategy:

\begin{lstlisting}[basicstyle=\small\ttfamily,breaklines=true,frame=single]
You are a research assistant with access to a
knowledge graph and document search.
Your goal is to answer the user's question as
accurately and completely as possible.

STRATEGY:
1. First, use search_documents to find relevant
   documents.
2. Examine the results for linked entity URLs
   (especially data.wordlift.io URLs).
3. If the question requires information about
   related entities (provider, offers, etc.),
   use follow_entity_link to fetch their data.
4. If you need to discover entities not directly
   linked, use search_knowledge_graph.
5. Synthesize all gathered information into a
   comprehensive answer.
6. Cite your sources by document ID or entity
   URL.

RULES:
- Only state facts you found in the retrieved
  data. Do not hallucinate.
- Follow at most {max_hops} link hops.
- If you cannot find sufficient information,
  say so explicitly.
\end{lstlisting}

\subsection{LLM Judge Prompts}

Three evaluation prompts are used by the independent judge model
(Gemini 3 Flash Preview). All judge calls request JSON output via
\texttt{response\_mime\_type="application/json"}.

\paragraph{Accuracy (1--5):}
\begin{lstlisting}[basicstyle=\small\ttfamily,breaklines=true,frame=single]
You are an evaluation judge. Given a question,
a ground-truth answer, and a candidate answer
produced by a system, rate the candidate's
factual accuracy.

Question: {question}
Ground Truth Answer: {ground_truth}
Candidate Answer: {candidate}

Rate the accuracy on a scale of 1-5:
  1 = Completely wrong or irrelevant
  2 = Mostly wrong, with minor correct elements
  3 = Partially correct, missing key facts
  4 = Mostly correct, minor inaccuracies
  5 = Fully correct and accurate

Respond in JSON format ONLY:
{"score": <1-5>, "reasoning": "brief explanation"}
\end{lstlisting}

\paragraph{Completeness (1--5):}
\begin{lstlisting}[basicstyle=\small\ttfamily,breaklines=true,frame=single]
You are an evaluation judge. Given a question,
a ground-truth answer containing key facts, and
a candidate answer, rate the completeness of the
candidate.

Question: {question}
Ground Truth (contains key facts the answer
should cover): {ground_truth}
Candidate Answer: {candidate}

Rate completeness on a scale of 1-5:
  1 = Covers none of the key facts
  2 = Covers ~25% of key facts
  3 = Covers ~50% of key facts
  4 = Covers ~75% of key facts
  5 = Covers all key facts

Respond in JSON format ONLY:
{"score": <1-5>, "facts_covered": [...],
 "facts_missing": [...],
 "reasoning": "brief explanation"}
\end{lstlisting}

\paragraph{Grounding (binary):}
\begin{lstlisting}[basicstyle=\small\ttfamily,breaklines=true,frame=single]
You are an evaluation judge. Given a candidate
answer and the source documents it was based on,
determine what fraction of claims in the answer
are traceable to the source documents.

Candidate Answer: {candidate}
Source Documents: {sources}

For each claim in the answer, determine if it
is supported by the source documents.

Respond in JSON format ONLY:
{"grounding_score": <0.0-1.0>,
 "total_claims": <int>,
 "grounded_claims": <int>,
 "ungrounded_claims": ["claim1", "claim2"]}
\end{lstlisting}

\subsection{Enhanced Entity Page Template}

The enhanced entity page template (Jinja2) used for condition C3 and C6
documents. Key features: Schema.org type breadcrumbs, embedded JSON-LD,
visible linked entity navigation with content negotiation instructions,
and \texttt{llms.txt}-style agent instructions.

\begin{lstlisting}[basicstyle=\small\ttfamily,breaklines=true,frame=single,
  language=HTML]
<!DOCTYPE html>
<html lang="en"
  prefix="schema: http://schema.org/">
<head>
  <title>{{ entity_name }}
    -- {{ domain_name }}</title>
  <script type="application/ld+json">
    {{ jsonld_data }}
  </script>
</head>
<body vocab="http://schema.org/"
  typeof="{{ entity_types }}">

  <!-- Type breadcrumbs -->
  <nav class="breadcrumb">
    Thing > CreativeWork > {{ entity_types }}
  </nav>

  <h1 property="name">{{ entity_name }}</h1>
  <div class="entity-type">
    Type: schema:{{ entity_types }}
  </div>

  <section><h2>Description</h2>
    <p property="description">
      {{ entity_description }}
    </p>
  </section>

  <!-- Linked entity navigation -->
  <section class="linked-entities">
    <h2>Related Entities</h2>
    <p>Each URI supports content negotiation
      -- append .json for JSON-LD, .ttl for
      Turtle, or .html for a human-readable
      view.</p>
    <ul>
    {% for le in linked_entities %}
      <li>
        <span class="relation">
          {{ le.relation }}:
        </span>
        <a href="{{ le.url }}">{{ le.url }}</a>
      </li>
    {% endfor %}
    </ul>
  </section>

  <!-- Agent instructions (llms.txt style) -->
  <section class="agent-instructions">
    <h2>Agent Instructions</h2>
    <pre>{{ llms_instructions }}</pre>
  </section>

</body></html>
\end{lstlisting}

\subsection{Enhanced+ Entity Page Template (C6+)}
\label{app:enhanced_plus}

The Enhanced+ entity page template extends the base enhanced format
(Appendix~\ref{app:prompts}) with several additional navigational
affordances designed to maximize agentic discoverability:

\begin{itemize}
    \item \textbf{Wikidata-style statements table}: Structured
          property--value pairs displayed in a scannable table format,
          with property names linking to their Schema.org definitions
          and values linking to related entities when applicable.
    \item \textbf{Named linked entities}: Related entity links include
          human-readable names (e.g., ``BlackBriar USA'') instead of
          raw URIs, reducing cognitive load for both agents and humans.
    \item \textbf{Sitelinks section}: Quick-access links to the
          entity's canonical web page and machine-readable serializations
          (JSON-LD, Turtle, RDF/XML).
    \item \textbf{Visual styling}: Card-based layout with distinct
          sections for description, statements, linked entities,
          sitelinks, and agent instructions, improving visual
          hierarchy and scannability.
    \item \textbf{Entity URI display}: Explicit display of the
          entity's dereferenceable URI in the header, making the
          knowledge graph node identity visible.
\end{itemize}

Key structural differences from the base enhanced template:
\begin{lstlisting}[basicstyle=\small\ttfamily,breaklines=true,frame=single,
  language=HTML]
<!-- Statements (Wikidata-style) -->
<table class="statements-table">
  <tr>
    <td class="prop-name">
      <a href="https://schema.org/offers">
        offers</a></td>
    <td><a href="https://data.wordlift.io/
      wl172055/offer/...">$49.99</a></td>
  </tr>
</table>

<!-- Named linked entities -->
<li>
  <span class="relation">brand:</span>
  <a href="/.../blackbriar-usa.html"
     rel="brand">BlackBriar USA</a>
</li>

<!-- Sitelinks -->
<a href="{{ canonical_url }}">Visit Web Page</a>
<a href="{{ entity_url }}.json">JSON-LD</a>
<a href="{{ entity_url }}.ttl">Turtle</a>
\end{lstlisting}

The Enhanced+ template achieves the highest scores in our evaluation
(accuracy: 4.85/5, completeness: 4.55/5), though the improvement over
the base enhanced format (C6) is not statistically significant
($\Delta = +0.06$, $p_\text{adj} = 1.0$), suggesting that the core
enhanced page design captures most of the benefit. The full template
source is available in the repository at
\texttt{templates/enhanced\_entity\_plus.html}.

\end{document}